%% file: main.tex
\documentclass[aps,pre,reprint,superscriptaddress,amsmath,amssymb]{revtex4-2}

\usepackage{graphicx}
\usepackage{txfonts}
\usepackage[hypertexnames=false]{hyperref}
\usepackage{braket}
\usepackage{bm}
\usepackage{enumitem}

\begin{document}

\title{Beyond local detailed balance: Microscopic rates reshape nonequilibrium phase behavior}

\author{Takahiro Kanazawa}
\email{kanazawa@uchicago.edu}
\affiliation{Department of Physics, The University of Tokyo, 7-3-1 Hongo, Bunkyo-ku, Tokyo 113-0033, Japan}
\affiliation{RIKEN Pioneering Research Institute, 2-1 Hirosawa, Wako 351-0198, Japan}
\affiliation{Department of Physics, The University of Chicago, 5720 South Ellis Avenue, Chicago, Illinois 60637, USA}
\author{Kyogo Kawaguchi}
\email{kyogo.kawaguchi@riken.jp}
\affiliation{Department of Physics, The University of Tokyo, 7-3-1 Hongo, Bunkyo-ku, Tokyo 113-0033, Japan}
\affiliation{RIKEN Pioneering Research Institute, 2-1 Hirosawa, Wako 351-0198, Japan}
\affiliation{Institute for Physics of Intelligence, The University of Tokyo, 7-3-1 Hongo, Bunkyo-ku, Tokyo 113-0033, Japan}
\author{Kyosuke Adachi}
\email{kyosuke.adachi@riken.jp}
\affiliation{RIKEN Pioneering Research Institute, 2-1 Hirosawa, Wako 351-0198, Japan}
\affiliation{RIKEN Center for Interdisciplinary Theoretical and Mathematical Sciences, 2-1 Hirosawa, Wako 351-0198, Japan}
\affiliation{RIKEN Center for Biosystems Dynamics Research, 2-2-3 Minatojima-minamimachi, Chuo-ku, Kobe 650-0047, Japan}

\date{\today}

\begin{abstract}
Local detailed balance (LDB) is a central guiding principle for modeling nonequilibrium stochastic dynamics, yet it only constrains the ratio of forward and backward transition rates and does not fix the steady state.
Although the functional form of rates under the same LDB has been shown to affect steady-state correlations and transition temperatures, whether it can qualitatively reshape phase behavior in strongly interacting systems, such as the morphology and stability of phase-separated patterns, remains unclear.
Here, for a two-dimensional driven lattice gas with attractive nearest-neighbor interactions, we consider hopping rates with a parameter that preserves the same LDB but tunes asymmetry along the driving force.
We find that this parameter controls qualitative phase behavior: in the homogeneous phase, it reverses the sign of the structure-factor discontinuity and hence the anisotropy in long-range density correlations; in the phase-separated regime, it switches the orientation of anisotropic patterns and their long-time stability.
Both effects are coherently captured by an approximate fluctuating hydrodynamic equation.
The results demonstrate that, in contrast to equilibrium systems, qualitative features of nonequilibrium phases are controlled by the specific choice of dynamical rules even under the same LDB.
\end{abstract}

\maketitle

\begin{figure*}[t]
    \centering
    \includegraphics[scale=1]{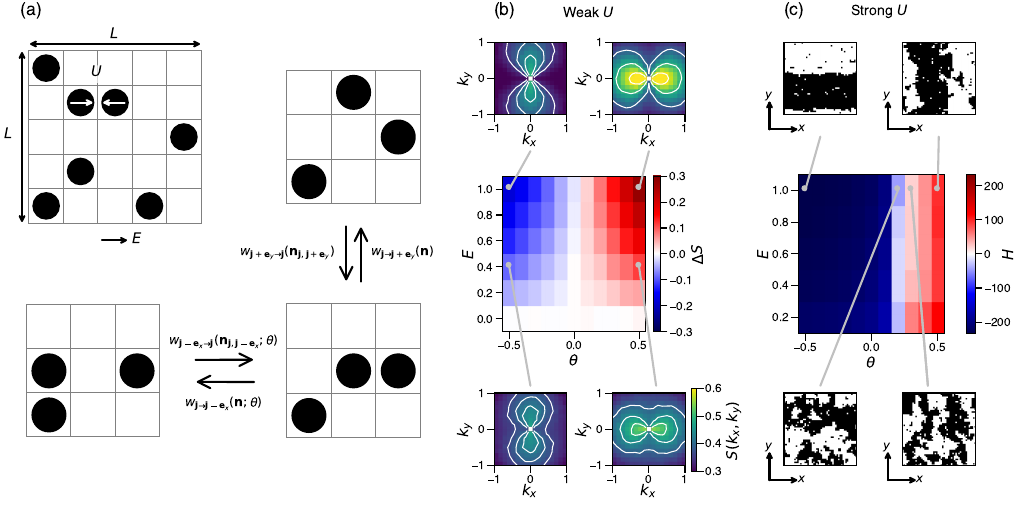}
    \caption{Driven lattice gas parameterized by $\theta$ and qualitative steady-state phase diagrams.
    (a) We consider particles with onsite exclusion, nearest-neighbor attraction $U$, and external field $E$ along the $x$ axis.
    The parameter $\theta$ controls asymmetry in the hopping rate along the field direction, while preserving the same LDB.
    (b) Phase diagram for homogeneous states with weak attractive interaction ($U = 0.5$), represented by structure factor discontinuity $\Delta S$.
    We also plot heatmaps of anisotropic structure factor $S(\bm{k})$ at indicated parameter sets.
    (c) Phase diagram for phase-separated states with strong attractive interaction ($U = 2$), represented by the peak height $H$ of structure factor, signed by the peak direction.
    Positive $H$ (red) and negative $H$ (blue) indicate that the peak appears on the $k_x$ and $k_y$ axes, respectively.
    We also show typical configurations.}
    \label{fig_model}
\end{figure*}

\textit{Introduction}---%
Nonequilibrium many-body systems exhibit diverse collective phenomena with no equilibrium counterpart, including flocking and motility-induced phase separation in active matter~\cite{Marchetti2013, Chate2020, CatesTailleur2015, CatesNardini2025}, and generic long-range correlations and anisotropic ordering in externally driven systems~\cite{SchmittmannZia1995, Marro1999, Zia2010}.
Driven lattice gases~\cite{Katz1983, Katz1984} provide a paradigmatic setting in which power-law density correlations and drive-dependent phase separation can be observed within a single model~\cite{SchmittmannZia1995, Marro1999, Zia2010}.
Classifying such nonequilibrium phase behavior in terms of microscopic ingredients, such as the particle interactions, external driving field, and form of dynamical rates, remains an open problem of nonequilibrium statistical mechanics.

Local detailed balance (LDB) provides a general framework for constructing physically consistent stochastic models of systems coupled to thermal, chemical, or mechanical reservoirs~\cite{Maes2021}.
It fixes the ratio of forward and backward transition rates in terms of entropy production and underlies a large family of exact nonequilibrium relations, including fluctuation theorems and thermodynamic uncertainty relations~\cite{Seifert2012, BaratoSeifert2015}.
However, LDB leaves the individual rates and therefore the steady-state distribution undetermined: different dynamical rules compatible with the same LDB generally lead to different steady states, in contrast to equilibrium processes, in which global detailed balance uniquely selects the Boltzmann distribution~\cite{vanKampen2007, Tauber2014}.

The consequences of this freedom of rates have been explored in several systems.
For dynamical models of molecular motors with only a few degrees of freedom, distinct rate parameterizations consistent with the same LDB yield different statistical properties, such as heat dissipation~\cite{Kawaguchi2014, Wang2018} and force-velocity curves~\cite{Blackwell2019}.
For weakly interacting driven particles, the presence or absence of long-range density correlations depends on the form chosen for the hopping rate~\cite{Tasaki2004, LefevereTasaki2005}.
For the driven lattice gas, the steady state, including its structure factor, has been shown to depend on the choice among Metropolis, Glauber, and heat-bath update rules, the first two of which obey the same LDB ratio~\cite{Kwak2004}.
For the asymmetric simple exclusion process, even a spatially local change of the rates at a single defect bond, which leaves the LDB ratio intact, can induce macroscopic segregation~\cite{Janowsky1992}.
In contrast, it remains largely unexplored whether a continuous, spatially uniform deformation of the rates under the same LDB can qualitatively reshape many-body behavior, such as the morphology or stability of phase-separated patterns, in strongly interacting nonequilibrium systems.

Here, we address this question using a two-dimensional driven lattice gas with hard-core exclusion and nearest-neighbor attraction under an external field.
We introduce hopping rates parameterized by an asymmetry parameter $\theta$, which preserves the same LDB while smoothly deforming the dynamical rule; the standard exponential rate corresponds to $\theta = 0$~\cite{vanBeijeren1984, Krug1986, Valles1986, Tasaki2004, LefevereTasaki2005}.
For weak interactions, the sign of the structure-factor discontinuity at the origin, which reflects the anisotropy of the long-range density correlation, reverses as $\theta$ crosses zero, rotating the structure-factor pattern by $90^\circ$.
For strong interactions, the same parameter controls the orientation of phase-separated clusters and their dynamical stability: vertically aligned patterns are broken and regenerated repeatedly, whereas horizontally aligned patterns remain stable.
Both phenomena are coherently captured by an approximate fluctuating hydrodynamic equation with microscopically derived $\theta$-dependent coefficients.
These results emphasize that, in contrast to equilibrium phases, not only LDB but also the specific choice of dynamical rule is key to modeling nonequilibrium phases.

\textit{Driven lattice gas with parameterized hopping rates}---%
We consider a lattice gas model with total particle number $N$ on a square lattice of size $(L, L)$ and periodic boundary conditions [Fig.~\ref{fig_model}(a)].
The lattice site is specified by $\bm{j} := j_x \bm{e}_x + j_y \bm{e}_y =: (j_x, j_y)$, where $\bm{e}_a$ is the unit vector in the $a$ direction, $j_a \in \{ 1, 2, \cdots, L \}$, and $a \in \{ x, y \}$.
The particle configuration is specified by $\bm{n} := \{ n_{\bm{j}} \}$, where $n_{\bm{j}} \in \{ 0, 1 \}$ is the occupation of site $\bm{j}$.

We assume that particles can stochastically hop to the neighboring sites with hard-core exclusion and nearest-neighbor attractive interactions of strength $U \geq 0$ under an external field of strength $E \geq 0$.
To implement this setup in a Markov process, we employ the LDB as a constraint on the ratio of forward and backward hopping rates:
\begin{equation}
    \begin{aligned}
        w_{\bm{j} \to \bm{j} + \bm{e}_x} (\bm{n}) / w_{\bm{j} + \bm{e}_x \to \bm{j}} (\bm{n}_{\bm{j}, \bm{j} + \bm{e}_x}) & = \exp [U \Delta N^\mathrm{nn}_{\bm{j} \to \bm{j} + \bm{e}_x} (\bm{n}) + E] \\
        w_{\bm{j} \to \bm{j} + \bm{e}_y} (\bm{n}) / w_{\bm{j} + \bm{e}_y \to \bm{j}} (\bm{n}_{\bm{j}, \bm{j} + \bm{e}_y}) & = \exp [U \Delta N^\mathrm{nn}_{\bm{j} \to \bm{j} + \bm{e}_y} (\bm{n})].
        \label{eq_rate_ratio}
    \end{aligned}
\end{equation}
Here, $w_{\bm{j} \to \bm{j}'} (\bm{n})$ is the hopping rate of a particle from site $\bm{j}$ to an empty site $\bm{j}'$ for configuration $\bm{n}$, $\Delta N^\mathrm{nn}_{\bm{j} \to \bm{j}'} (\bm{n})$ is the change in the number of neighboring particles due to hopping, $\bm{n}_{\bm{j}, \bm{j}'}$ is the configuration after moving a particle from site $\bm{j}$ to $\bm{j}'$ starting from configuration $\bm{n}$, and temperature is used as the unit of energy.
A nonzero $E$ breaks global detailed balance, making the process nonequilibrium.

To examine how the choice of transition rates can affect phase behavior, we consider the following rates parameterized by $\theta \in \mathbb{R}$ [Fig.~\ref{fig_model}(a)]:
\begin{equation}
    \begin{aligned}
        w_{\bm{j} \to \bm{j} + \bm{e}_x} (\bm{n}; \theta) & = \exp [(1 / 2 + \theta) (U \Delta N^\mathrm{nn}_{\bm{j} \to \bm{j} + \bm{e}_x} (\bm{n}) + E)] \\
        w_{\bm{j} \to \bm{j} - \bm{e}_x} (\bm{n}; \theta) & = \exp [(1 / 2 - \theta) (U \Delta N^\mathrm{nn}_{\bm{j} \to \bm{j} - \bm{e}_x} (\bm{n}) - E)] \\
        w_{\bm{j} \to \bm{j} + \bm{e}_y} (\bm{n}) & = \exp [(1 / 2) U \Delta N^\mathrm{nn}_{\bm{j} \to \bm{j} + \bm{e}_y} (\bm{n})] \\
        w_{\bm{j} \to \bm{j} - \bm{e}_y} (\bm{n}) & = \exp [(1 / 2) U \Delta N^\mathrm{nn}_{\bm{j} \to \bm{j} - \bm{e}_y} (\bm{n})].
        \label{eq_rate}
    \end{aligned}
\end{equation}
The parameter $\theta$ controls asymmetry in the hopping rate along the driving field.
Similar rates with the exponential form with $\theta = 0$ have been considered in previous studies~\cite{vanBeijeren1984, Krug1986, Valles1986, Tasaki2004, LefevereTasaki2005}.

The parameter $\theta$ can be interpreted as a kinetic parameter as follows.
Writing $X := U \Delta N^\mathrm{nn}_{\bm{j} \to \bm{j} + \bm{e}_x} (\bm{n}) + E$ for the entropy production associated with a hop in the field direction, we can decompose the rates for this hop and its reverse [Eq.~\eqref{eq_rate}] as $w = e^{\pm X / 2} e^{\theta X}$.
The time-antisymmetric factor $e^{\pm X / 2}$ is fixed by the LDB condition~\eqref{eq_rate_ratio}, while the time-symmetric factor $e^{\theta X}$ controls the kinetic reactivity of the transition pair, and thereby affects the dynamical activity tuned by $\theta$.
The value $1 / 2 + \theta$ plays the role of a load-distribution factor, here generalized to include the nearest-neighbor interactions; load-distribution factors have been estimated for biomolecular motors~\cite{Fisher2001}.

We stress that the LDB condition~\eqref{eq_rate_ratio} is kept intact regardless of the choice of $\theta$.
In the equilibrium limit where $E = 0$, LDB leads to global detailed balance, and the steady-state distribution $P^\mathrm{ss} (\bm{n})$ is the Boltzmann distribution independent of $\theta$~\cite{vanKampen2007, Tauber2014}.
In contrast, for nonequilibrium systems with $E > 0$, the choice of $\theta$ affects the qualitative features of $P^\mathrm{ss} (\bm{n})$, as demonstrated in the subsequent sections.

\begin{figure}[t]
    \centering
    \includegraphics[scale=1]{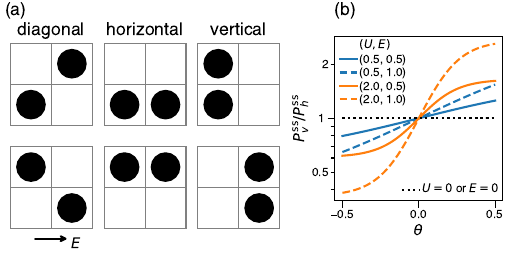}
    \caption{$\theta$-dependent configurational biases in the two-particle system with $L = 2$. (a) Diagonally, horizontally, and vertically aligned configurations. (b) Ratio of the steady-state probabilities for vertically and horizontally aligned configurations, $P_v^{\mathrm{ss}}/P_h^{\mathrm{ss}}$ [Eq.~\eqref{eq_two_particle_anisotropy}]. When $U=0$ or $E=0$, $P_v^{\mathrm{ss}}/P_h^{\mathrm{ss}}=1$ for any $\theta$.}
    \label{fig_two_particle}
\end{figure}

\textit{Analytical result for two-particle steady state}---%
To grasp how the $\theta$-dependent behavior can emerge in systems with identical LDB, we consider two particles on a lattice with $L = 2$.
Using the rates~\eqref{eq_rate} and the master equation for the steady state, we obtain the following relations:
\begin{align}
    e^{U + \theta E} \cosh (2 \theta U + E / 2) P^\mathrm{ss}_d & = e^{-U + \theta E} \cosh (-2 \theta U + E / 2) P^\mathrm{ss}_v \nonumber \\
    e^U P^\mathrm{ss}_d & = e^{-U} P^\mathrm{ss}_h, \label{eq_two_particle_steady_relation}
\end{align}
where $P^\mathrm{ss}_d$, $P^\mathrm{ss}_h$, and $P^\mathrm{ss}_v$ represent the steady-state probabilities of taking diagonally, horizontally, and vertically aligned particle configurations, respectively [Fig.~\ref{fig_two_particle}(a)]; for example, $P^\mathrm{ss}_v = P^\mathrm{ss} (n_{(1, 1)} = 1, n_{(1, 2)} = 1, n_{(2, 1)} = 0, n_{(2, 2)} = 0) + P^\mathrm{ss} (n_{(1, 1)} = 0, n_{(1, 2)} = 0, n_{(2, 1)} = 1, n_{(2, 2)} = 1)$.

From Eq.~\eqref{eq_two_particle_steady_relation}, we obtain
\begin{equation}
    \frac{P^\mathrm{ss}_v}{P^\mathrm{ss}_h} = \frac{1 + \tanh (2 \theta U) \tanh (E / 2)}{1 - \tanh (2 \theta U) \tanh (E / 2)}.
    \label{eq_two_particle_anisotropy}
\end{equation}
This suggests that vertically and horizontally aligned particle configurations are relatively favorable for $\theta > 0$ and $\theta < 0$, respectively, for attractive interactions ($U > 0$) with driving in the $x$ direction ($E > 0$).
In Fig.~\ref{fig_two_particle}(b), we plot the $\theta$-dependent behavior of $P^\mathrm{ss}_v / P^\mathrm{ss}_h$ for different values of $U$ and $E$.

From Eq.~\eqref{eq_two_particle_anisotropy}, we see that both nonzero interaction $U$ and nonzero driving $E$ are essential to the $\theta$-dependent steady state $P^\mathrm{ss} (\bm{n})$.
This property actually holds for many-particle systems in general.
First, for $E = 0$, $P^\mathrm{ss} (\bm{n})$ is the Boltzmann distribution and is independent of $\theta$, as mentioned above.
Second, for $U = 0$, $P^\mathrm{ss} (\bm{n})$ is independent of $E$ or $\theta$ since the model is equivalent to the two-dimensional asymmetric simple exclusion process with periodic boundary conditions~\cite{Kipnis2010}, as known also for one-dimensional systems~\cite{Spohn1991, Blythe2007}.
In the following, we consider many-particle systems with $E > 0$ and $U > 0$ and illustrate how $\theta$-dependent phase behavior appears.

\textit{$\theta$-dependent anisotropy in long-range correlation}---%
In several driven lattice gas models with weak interactions, breaking detailed balance with spatial anisotropy leads to a long-range density correlation and an associated singularity in the structure factor $S (\bm{k})$ in the steady state~\cite{SchmittmannZia1995, Zia2010}.
Here, we examine how $\theta$ can control the density correlation by considering a system with $L = 50$, $N = 1250$ (i.e., half-filling), and $U = 0.5$.

We conducted Monte Carlo (MC) simulations to sample the steady-state configurations for several values of $\theta$ and $E$ (see Appendix~\ref{app_simulation} for simulation details; simulation codes are available at Ref.~\cite{code} and can be run on Google Colab to generate videos).
We confirmed that the steady state is a homogeneous phase for any parameter set used here since the interaction $U$ is weak enough.
To quantify the long-range correlation in the Fourier space, we measured the signed discontinuity in the steady-state structure factor, $\Delta S := \lim_{k_x \to 0} S(k_x, k_y = 0) - \lim_{k_y \to 0} S(k_x = 0, k_y)$.
Here, $k_a := 2 \pi n_a / L \in [-\pi, \pi)$ with $n_a \in \mathbb{Z}$ ($a \in \{ x, y \}$), and the structure factor is defined as $S(\bm{k}) := \braket{|\sum_{\bm{j}} e^{-i \bm{k} \cdot \bm{j}} n_{\bm{j}}|^2} / L^2$ for $(k_x, k_y) \neq (0, 0)$, where $\braket{\cdots}$ represents the steady-state average.
For $\Delta S \neq 0$, the density correlation should asymptotically follow $C (x, y) \propto -\Delta S (x^2 - y^2) / (x^2 + y^2)^2$ in a rescaled coordinate~\cite{SchmittmannZia1995}.

\begin{figure}[t]
    \centering
    \includegraphics[scale=1]{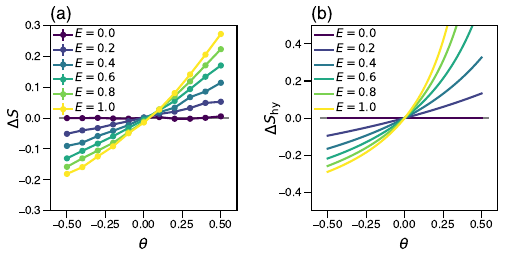}
    \caption{$\theta$-dependent anisotropy in long-range density correlation.
    (a) Discontinuity in the structure factor $\Delta S$ as a function of $\theta$ for various $E$, obtained numerically from lattice gas simulations.
    (b) Counterpart of (a), obtained analytically from fluctuating hydrodynamics.}
    \label{fig_structure_factor}
\end{figure}

In Fig.~\ref{fig_model}(b), we plot the heatmap of $\Delta S$ in the $\theta$--$E$ plane, where red and blue indicate positive and negative values of $\Delta S$, respectively.
We see that $\Delta S = 0$ at $E = 0$, suggesting that no long-range correlation appears in the equilibrium limit.
In contrast, for $E > 0$, we find that $\Delta S$ changes its sign from negative (blue) to positive (red) at $\theta \approx 0$ when $\theta$ increases from $-0.5$ to $0.5$.
As seen from the example heatmaps of $S (\bm{k})$ at four parameter sets in Fig.~\ref{fig_model}(b), the anisotropic pattern in $S (\bm{k})$ rotates by $90^\circ$ when crossing $\theta \approx 0$.
This illustrates that $\theta$ controls the long-range correlation property in the homogeneous phase.

We also analytically obtained the $\theta$-dependent density correlation by extending the perturbation theory developed in Ref.~\cite{LefevereTasaki2005} (see Appendix~\ref{app_high_temp_exp} for the derivation).
At the leading order in the high-temperature expansion, we find that the two-body part of the nonequilibrium effective interaction exhibits a power-law decay proportional to $EU\theta$.
This can be regarded as the origin of the long-range correlation with $\sim EU\theta/r^2$~\cite{LefevereTasaki2005}.
This result suggests that nonzero $\theta$ is essential to the emergence of long-range correlation at the leading order of $U$ and $E$.
% Furthermore, the sign of $\theta$ changes the direction of anisotropic pattern(?).

\begin{figure*}[t]
    \centering
    \includegraphics[scale=1]{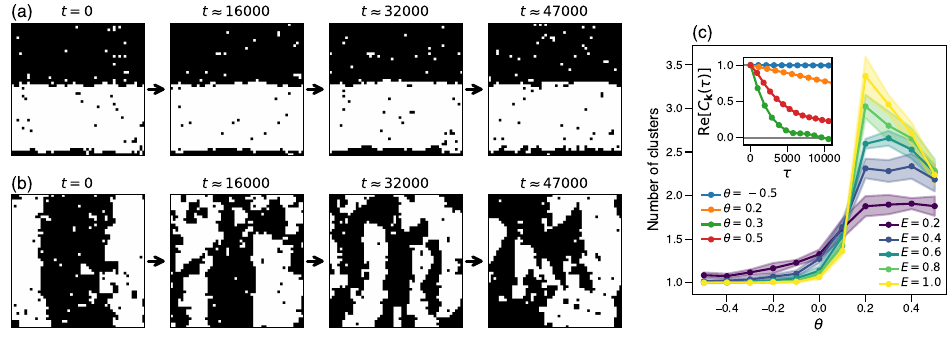}
    \caption{$\theta$-dependent morphology and stability of phase separation.
    (a), (b) Snapshots for typical dynamics at (a) $\theta = -0.5$ (horizontally aligned stable pattern) and (b) $\theta = 0.5$ (vertically aligned unstable pattern).
    (c) Total number of clusters in the steady state as a function of $\theta$.
    The brightness represents the strength of $E$, and the shadow for each curve indicates the standard deviation.
    Inset: for $E = 1$, we plot the autocorrelation of large-scale patterns, $\mathrm{Re} [C_{\bm{k}}(\tau)]$.}
    \label{fig_phase_separation}
\end{figure*}

To explain the $\theta$ dependence of $S (\bm{k})$ from a hydrodynamic viewpoint, we derived an approximate fluctuating hydrodynamic equation~\cite{Andreanov2006, Lefevre2007} from our lattice gas model (see Appendix~\ref{app_hydro} for the derivation).
Assuming that the phase is sufficiently homogeneous and nonlinearity in density fluctuation is negligible, we obtain a linear hydrodynamic equation:
\begin{equation}
    \partial_t \phi = A_x (U, E, \theta) {\partial_x}^2 \phi + A_y (U) {\partial_y}^2 \phi + \xi,
    \label{eq_linear_fluct_hydro}
\end{equation}
where $\phi (\bm{r}, t)$ is the field of density fluctuation around mean density ($= 1 / 2$), and $\xi (\bm{r}, t)$ is a Gaussian noise field satisfying $\braket{\xi (\bm{r}, t)} = 0$ and $\braket{\xi (\bm{r}, t) \xi (\bm{r}', t')} = -2 [B_x (E, \theta) {\partial_x}^2 + B_y {\partial_y}^2] \delta (\bm{r} - \bm{r}') \delta (t - t')$.
The form of Eq.~\eqref{eq_linear_fluct_hydro} itself is the linear Langevin equation known for driven diffusive systems~\cite{SchmittmannZia1995, Zia2010}; the new ingredient here is the microscopically derived dependence of the coefficients on $\theta$ as well as on $U$ and $E$.
The relation between the coefficients and lattice gas model parameters is given by $A_x = e^{\theta E} \cosh (E / 2) \{ 1 - (5 U / 4) [1 + 2 \theta \tanh (E / 2)] \}$, $A_y = 1 - 5 U / 4$, $B_x = e^{\theta E} \cosh (E / 2) / 4$, and $B_y = 1 / 4$.

From Eq.~\eqref{eq_linear_fluct_hydro}, we obtain the structure factor $S_\mathrm{hy} (\bm{k})$ and its discontinuity at the origin, $\Delta S_\mathrm{hy} := \lim_{k_x \to 0} S_\mathrm{hy} (k_x, k_y = 0) - \lim_{k_y \to 0} S_\mathrm{hy} (k_x = 0, k_y)$:
\begin{equation}
    \Delta S_\mathrm{hy} = \frac{1}{4 - 5 U [1 + 2 \theta \tanh (E / 2)]} - \frac{1}{4 - 5 U}.
    \label{eq_hydro_structure_factor}
\end{equation}
This functional form suggests that the driving force $E > 0$ induces the discontinuity, the sign of which is determined by that of $\theta$.
In Fig.~\ref{fig_structure_factor}, we compare $\Delta S$ observed in lattice gas simulations [Fig.~\ref{fig_structure_factor}(a)] and $\Delta S_\mathrm{hy}$ predicted from the approximate hydrodynamics [Fig.~\ref{fig_structure_factor}(b)].
Apart from the quantitative difference, we find a similar $E$ and $\theta$ dependence, particularly the sign change when crossing $\theta \approx 0$ for $E > 0$.
This indicates that the fluctuating hydrodynamic equation~\eqref{eq_linear_fluct_hydro} captures the essential nonequilibrium phase behavior of the driven lattice gas model parameterized by $\theta$.

\textit{$\theta$-dependent morphology and stability of phase separation}---%
We consider the case in which the attractive interaction $U$ is strong enough for phase separation to occur.
Specifically, we fix $L = 50$, $N = 1250$ (i.e., half-filling), and $U = 2$, with varying $\theta$ and $E$.

In Fig.~\ref{fig_model}(c), we plot a qualitative steady-state phase diagram, shown with typical particle configurations.
The heatmap represents the signed peak height of the structure factor, $H := s \max_{\bm{k}} S(\bm{k})$, where $s = 1$ if $\arg \max_{\bm{k}} S(\bm{k})$ is on the $k_x$ axis and $s = -1$ if $\arg \max_{\bm{k}} S(\bm{k})$ is on the $k_y$ axis.
The sign of $H$ changes depending on $\theta$ for $E > 0$, which suggests that the morphology of phase separation (i.e., horizontal or vertical alignment) is switched by $\theta$, as demonstrated by the typical configurations in Fig.~\ref{fig_model}(c).

The observed $\theta$-dependent anisotropy in phase separation can be qualitatively explained by the linear hydrodynamic equation~\eqref{eq_linear_fluct_hydro} near the stability limit of the homogeneous state.
The deterministic part of the density fluctuation follows
\begin{align}
    & \partial_t \phi_{\bm{k}} = -[e^{\theta E} \cosh (E / 2) (1 - 5 U_x^\mathrm{eff} / 4) {k_x}^2 + (1 - 5 U / 4) {k_y}^2] \phi_{\bm{k}}, \nonumber \\
    & U_x^\mathrm{eff} (E, \theta) := U [1 + 2 \theta \tanh (E / 2)],
    \label{eq_linear_hydro_Fourier}
\end{align}
where $\phi_{\bm{k}} (t) := \int d \bm{r} e^{-i \bm{k} \cdot \bm{r}} \phi (\bm{r}, t)$, and $U_x^\mathrm{eff}$ is regarded as the effective attractive interaction along the $x$ axis.
The consequence of the $\theta$-dependent $U_x^\mathrm{eff}$ under $E > 0$ is interpreted as follows: a positive $\theta$ enhances $U_x^\mathrm{eff}$ and tends to destabilize the homogeneous state toward vertically aligned patterns, signaled by the coefficient of ${k_x}^2$ crossing zero, as $U$ increases; in contrast, negative $\theta$ weakens $U_x^\mathrm{eff}$ and results in a tendency toward horizontally aligned patterns as $U$ increases.
Note that a similar $\theta$-dependent anisotropy was found even at the two-particle level [Eq.~\eqref{eq_two_particle_anisotropy}].

Though the deterministic linear hydrodynamics captures the instability toward phase separation, steady-state properties are also influenced by nonlinearity and noise, potentially in a $\theta$-dependent way.
We tested the qualitative difference between horizontally and vertically aligned patterns by examining phase separation dynamics.
In Figs.~\ref{fig_phase_separation}(a) and \ref{fig_phase_separation}(b), we show typical steady-state dynamics for $\theta < 0$ and $\theta > 0$ cases, respectively.
We find that horizontally aligned patterns are stable [Fig.~\ref{fig_phase_separation}(a)], while vertically aligned patterns are broken and regenerated repeatedly [Fig.~\ref{fig_phase_separation}(b)].

More quantitatively, as shown in Fig.~\ref{fig_phase_separation}(c), we measured the $\theta$ and $E$ dependence of the total number of clusters, defined by connected components of particles with size larger than 9.
We find that the number of clusters remains close to unity for $\theta < 0$ (i.e., a single large cluster persists stably), whereas it increases for $\theta > 0$ (i.e., multiple small clusters coexist).
This contrast is more pronounced as the external field $E$ increases, and the peak observed at $\theta \approx 0.2$ is likely attributable to the rotational transition of the large-scale pattern, accompanied by its breakdown into smaller clusters, as shown in the snapshots of Fig.~\ref{fig_model}(c).
In the inset of Fig.~\ref{fig_phase_separation}(c), we also show the autocorrelation of large-scale patterns for $E = 1$, defined by $C_{\bm{k}}(\tau):=\langle \tilde{n}_{\bm{k}}(t) \tilde{n}_{\bm{k}}^*(t+\tau) \rangle / \langle |\tilde{n}_{\bm{k}}|^2 \rangle$ at the wavevector corresponding to the dominant pattern [$\bm{k}=(0, 2\pi/L)$ for $\theta=-0.5, 0.2$ and $\bm{k}=(2\pi/L, 0)$ for $\theta=0.3, 0.5$], where $\tilde{n}_{\bm{k}} := \sum_{\bm{j}} e^{-i \bm{k} \cdot \bm{j}} n_{\bm{j}}$.
We find that $\mathrm{Re} [C_{\bm{k}}(\tau)]$ decays faster for $\theta>0$, which quantifies the instability of vertical patterns observed in Fig.~\ref{fig_phase_separation}(b).
Since similar instabilities of bulk phase coexistence into multiple smaller phases have also been reported in active matter models~\cite{Tjhung2018, Shi2020, Adachi2026}, exploring the rate dependence of phase behavior in active systems is a natural direction for future work.

\textit{Discussion and conclusion}---%
We have shown that, in a two-dimensional driven lattice gas with attractive nearest-neighbor interactions, a one-parameter family of hopping rates consistent with the same LDB controls two types of nonequilibrium phase behaviors.
The parameter $\theta$ adjusts asymmetry between forward and backward hopping rates along the driving field $E$.
The sign and amplitude of $\theta$ control the anisotropy of long-range density correlations in the homogeneous phase and switch the orientation and long-time stability of patterns in the phase-separated regime.
Both effects are coherently captured by an approximate fluctuating hydrodynamic equation whose coefficients depend on $\theta$ and $E$.

Field-dependent transition rates have been measured in biomolecular motors, showing that forward and backward transitions need not respond symmetrically to an applied load~\cite{Schnitzer2000,Fisher2001,Kawaguchi2014}. By contrast, in the continuous overdamped Langevin limit, the path probability reduces to the standard Onsager--Machlup form: the microscopic field-splitting parameter is effectively fixed to $\theta=0$. This points to systems with an intrinsic discrete state space or lattice scale as the most relevant platforms for testing the present mechanism. Besides biomolecular motors, promising directions include open quantum (non-Hermitian) lattice models, where asymmetric hopping has been studied as a microscopic ingredient of nonequilibrium localization, topology, and interaction-driven phases~\cite{Hatano1996,adachi2022activity,takasan2024activity}.

The standard driven lattice gas with exponential or Metropolis rates exhibits anisotropic phase separation aligned with the drive~\cite{Katz1983, Katz1984, SchmittmannZia1995, Marro1999, Zia2010}, corresponding to the $\theta \leq 0$ regime in our model.
Increasing $\theta$ toward the positive side selects patterns aligned perpendicular to the drive, which are nonetheless disrupted and created repeatedly on long timescales.
A related orientational switch of patterns has been reported in two-temperature lattice gas models~\cite{Bassler1995} and two-species driven diffusive systems~\cite{Yu2022}, where it is induced by the change in driving forces; in contrast, our results show that the LDB-preserving reparameterization of transition rates is sufficient for a qualitative change in macroscopic patterns.
The dependence of steady states on the choice of the update rule has also been reported for the driven lattice gas~\cite{Kwak2004}; likewise, its critical temperature depends on the form of the rate function~\cite{Krug1986}.
To our knowledge, however, an orientational switch of phase separation induced solely by an LDB-preserving change of rates has not been demonstrated.
Meanwhile, when present, the long-range correlations in the homogeneous phase have the same form as those reported for driven lattice gases with other rate choices~\cite{SchmittmannZia1995, Zia2010, LefevereTasaki2005}, which suggests that the form of the correlation is universal; both its presence and its anisotropy, however, depend on the choice of rates, as shown here.
Our results caution against inferring the phase behavior of nonequilibrium many-body systems from a single convenient choice of transition rates.

\textit{Acknowledgments}---%
We thank R. Hamazaki and H. Nakano for fruitful discussions.
We also thank K. Orihara for helpful comments on the manuscript.
K.A. acknowledges support by the RIKEN Information systems division for the use of the Supercomputer HOKUSAI BigWaterfall2.
This work was supported by JSPS KAKENHI Grant Numbers JP26K00052 and JP26K17111 (to K.A.), as well as JP19H05275, JP23H00095, and JP25H01361 (to K.K.).
Claude Opus and Fable were used to improve readability.

\appendix

\section{Simulation of driven lattice gas model}
\label{app_simulation}

In numerical simulations of the driven lattice gas model [Eq.~\eqref{eq_rate}], we updated the particle configuration as follows:
\begin{enumerate}[label=(\arabic*)]
    \item We randomly choose a particle.
    \item The chosen particle (at site $\bm{j}$) hops to one of the empty adjacent sites (site $\bm{j}'$) with probability $w_{\bm{j} \to \bm{j}'} (\bm{n}) \Delta t$.
    \item We repeat procedures (1) and (2) $N$ times in a single MC step.
\end{enumerate}
We used a time step $\Delta t = [e^{(1/2 + \theta) (3 U + E)} + e^{(1/2 - \theta) (3 U - E)} + 2 e^{3U/2}]^{-1}$ to minimize the probability that hopping does not happen.

\begin{figure}[t]
    \centering
    \includegraphics[scale=1]{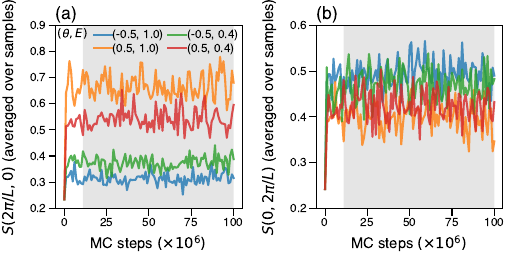}
    \caption{Relaxation dynamics of structure factors with weak attractive interactions. We plot the time evolution of the long-wavelength components of the structure factor $S(\bm{k})$ [$(k_x, k_y)=(2\pi/L, 0)$ for (a) and $(0, 2\pi/L)$ for (b)] averaged over independent samples for the parameter sets used in Fig.~\ref{fig_model}(b). The gray area indicates the time points used for averaging.}
    \label{fig_relaxation_weak}
\end{figure}

\begin{figure}[t]
    \centering
    \includegraphics[scale=1]{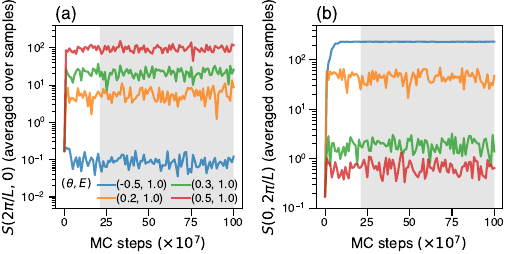}
    \caption{Relaxation dynamics of structure factors with strong attractive interactions. We plot the time evolution of the long-wavelength components of the structure factor $S(\bm{k})$ [$(k_x, k_y)=(2\pi/L, 0)$ for (a) and $(0, 2\pi/L)$ for (b)] averaged over independent samples for the parameter sets used in Fig.~\ref{fig_model}(c). The gray area indicates the time points used for averaging.}
    \label{fig_relaxation_strong}
\end{figure}

For Figs.~\ref{fig_model}(b) and \ref{fig_structure_factor}(a), we used $100$ independent samples at $90$ time points (taken every $10^6$ MC steps after relaxation of $1.1\times10^7$ MC steps, shown in Fig.~\ref{fig_relaxation_weak}); for Figs.~\ref{fig_model}(c) and \ref{fig_phase_separation}(c), we used $10$ independent samples at $80$ time points (taken every $10^7$ MC steps after relaxation of $2.1\times10^8$ MC steps, shown in Fig.~\ref{fig_relaxation_strong}).
To calculate the autocorrelation $C_{\bm{k}}(\tau)=\langle \tilde{n}_{\bm{k}}(t) \tilde{n}_{\bm{k}}^*(t+\tau) \rangle / \langle |\tilde{n}_{\bm{k}}|^2 \rangle$ plotted in Fig.~\ref{fig_phase_separation}(c), we used $10$ independent samples at $100$ different time points for averaging after relaxation of $2.1\times10^8$ MC steps.

\section{High-temperature expansions}
\label{app_high_temp_exp}

Following Ref.~\cite{LefevereTasaki2005}, we derive a high-temperature expansion for the steady state of the lattice gas model with nearest-neighbor interaction $U$, external field $E$, and the exponential dynamics parameterized by $\theta$ [Eq.~\eqref{eq_rate}].

% The probability measure of a configuration $\bm{n}$, $P(\bm{n})$, can be written as
% \begin{align}
%     P(\bm{n}) \propto \exp{[-\beta \tilde{H}(\bm{n})]} ,
% \end{align}
% where $\beta$ is the inverse temperature and $\tilde{H}(\bm{n})$ is a summable effective Hamiltonian.
% We write
% \begin{align}
%     \tilde{H}(\bm{n}) = H(\bm{n}) + \frac{\Psi(\bm{n})}{\beta}
% \end{align}
% with $H(\bm{n})$ and $\Psi(\bm{n})$ being the equilibrium Hamiltonian and the nonequilibrium effective interaction, respectively.

Let $\Lambda \subset \mathbb{Z}^2$ be the two-dimensional $L \times L$ lattice with periodic boundary conditions.
We denote the sites as $\bm{i} = (i_x, i_y) \in \Lambda$ and the configuration of the system as $\bm{n} = \{n_{\bm{i}}\}_{\bm{i}\in\Lambda}$.
The occupation variable $n_{\bm{i}} \in \{0, 1 \}$ is $0$ if the site $\bm{i}$ is empty and $1$ if it is occupied, where the total number of particles is $N=\sum_{\bm{i}\in\Lambda} n_{\bm{i}}$.
% Note that we consider the exclusion process, where each site can be occupied by at most one particle at a time.
% The interaction between particles is given by an Ising interaction.
The equilibrium Hamiltonian is given by
\begin{align}
    H(\bm{n}) = -\frac{U}{2} \sum_{\langle \bm{i},\bm{j} \rangle} n_{\bm{i}} n_{\bm{j}},
\end{align}
where $\sum_{\langle \bm{i},\bm{j} \rangle}$ denotes a sum over all pairs of sites with $|\bm{i}-\bm{j}|=1$, counting both $(\bm{i}, \bm{j})$ and $(\bm{j}, \bm{i})$.

The transition rate for a particle from $\bm{i}$ to $\bm{j}$ is given by
\begin{align}
    c(\bm{i} \to \bm{j}; \bm{n}) = \delta_{\bm{i} \to \bm{j}}^{(\bm{n})} \phi(\beta[H(\bm{n}^{\bm{i},\bm{j}})-H(\bm{n})+E(i_x-j_x)]) ,
    \label{eq:high_c}
\end{align}
where $\beta$ is the inverse temperature, $E$ is the external field in the $x$ direction, and $\delta_{\bm{i} \to \bm{j}}^{(\bm{n})} = n_{\bm{i}}(1-n_{\bm{j}})$ equals $1$ only when the transition is possible; otherwise, it is $0$.
We denote by $\bm{n}^{\bm{i},\bm{j}}$ the configuration in which $n_{\bm{i}}$ and $n_{\bm{j}}$ have been swapped in the original configuration $\bm{n}$.
For our transition rates, we have
\begin{align}
    \phi(h) &=
    \left\{ \,
    \begin{aligned}
    & \exp{\left[ - \left( \frac{1}{2} \pm \theta \right) h \right]} & (i_x-j_x=\mp1) \\
    & \exp{\left( -\frac{1}{2} h \right)} & (i_x-j_x=0)
    \end{aligned}
    \right. \\
    &=: \exp{\left[-\left(\frac{1}{2}+\tilde{\theta}_{\bm{i},\bm{j}}\right)h\right]} ,
    \label{eq:phi_h}
\end{align}
where $\tilde{\theta}_{\bm{i},\bm{j}}=\pm\theta$ for $i_x-j_x=\mp1$, respectively, and $\tilde{\theta}_{\bm{i},\bm{j}}=0$ for $i_x-j_x=0$.
From Eqs.~\eqref{eq:high_c}-\eqref{eq:phi_h}, the local detailed balance condition,
\begin{align}
    c(\bm{i} \to \bm{j}; \bm{n}) = e^{\beta[-H(\bm{n}^{\bm{i},\bm{j}})+H(\bm{n})-E(i_x-j_x)]} c(\bm{j} \to \bm{i}; \bm{n}^{\bm{i},\bm{j}}) ,
    \label{eq:high_c_relation}
\end{align}
is indeed satisfied.

The probability of configuration $\bm{n}$ at time $t$, $P_t(\bm{n})$, satisfies the master equation
\begin{align}
    \frac{d}{dt} P_t(\bm{n}) = \sum_{\langle \bm{i},\bm{j} \rangle} \left[ -P_t(\bm{n}) c(\bm{i} \to \bm{j}; \bm{n}) + P_t(\bm{n}^{\bm{i},\bm{j}}) c(\bm{j} \to \bm{i}; \bm{n}^{\bm{i},\bm{j}}) \right] .
\end{align}
By setting $dP_t(\bm{n})/dt=0$, we can obtain the steady-state distribution $P^{\mathrm{ss}}(\bm{n})$:
\begin{align}
    \sum_{\langle \bm{i},\bm{j} \rangle} \left[ P^{\mathrm{ss}}(\bm{n}) c(\bm{i} \to \bm{j}; \bm{n}) - P^{\mathrm{ss}}(\bm{n}^{\bm{i},\bm{j}}) c(\bm{j} \to \bm{i}; \bm{n}^{\bm{i},\bm{j}}) \right] = 0 .
    \label{eq:high_steady}
\end{align}
We assume that the steady-state distribution has a form
\begin{align}
    P^{\mathrm{ss}}(\bm{n}) = \frac{1}{Z(U,E)} \exp{[-\beta H(\bm{n}) - \tilde{\Psi}(\bm{n})]}
    \label{eq:high_prob}
\end{align}
where the effective interaction $\tilde{\Psi}(\bm{n})$ represents a nonequilibrium correction and $Z(U,E)$ is the normalization factor.
From Eqs.~\eqref{eq:high_steady} and \eqref{eq:high_prob}, we have
\begin{align}
    &\quad \sum_{\langle \bm{i}, \bm{j} \rangle} c(\bm{j} \to \bm{i}; \bm{n}^{\bm{i},\bm{j}}) e^{\beta [H(\bm{n})-H(\bm{n}^{\bm{i},\bm{j}})]} [1 - e^{\tilde{\Psi}(\bm{n})-\tilde{\Psi}(\bm{n}^{\bm{i},\bm{j}})}] \nonumber \\
    &= \sum_{\langle \bm{i},\bm{j} \rangle} \left[ c(\bm{j} \to \bm{i}; \bm{n}^{\bm{i},\bm{j}}) e^{\beta [H(\bm{n}) - H(\bm{n}^{\bm{i},\bm{j}})]} - c(\bm{i} \to \bm{j}; \bm{n}) \right] .
    \label{eq:high_start}
\end{align}

We expand $\tilde{\Psi}(\bm{n})$ in terms of $\beta$ and calculate the lowest-order contribution $\Psi(\bm{n})$, which is of order $O(\beta^2)$ as we will confirm below:
\begin{align}
    \tilde{\Psi}(\bm{n}) = \Psi(\bm{n}) + O(\beta^3) .
    \label{eq:high_Psi_expansion}
\end{align}
% We confirm that the leading order is indeed on the order of $\beta^2$ below.
First, $\tilde{\Psi}(\bm{n})=0$ when $\beta=0$.
Second, from Eqs.~\eqref{eq:high_c}, \eqref{eq:high_c_relation}, and \eqref{eq:high_Psi_expansion}, we find the left-hand side of Eq.~\eqref{eq:high_start} as
\begin{align}
    \sum_{\langle \bm{i}, \bm{j} \rangle} c(\bm{j} \to \bm{i}; \bm{n}^{\bm{i},\bm{j}}) e^{\beta [H(\bm{n})-H(\bm{n}^{\bm{i},\bm{j}})]} [1 - e^{\tilde{\Psi}(\bm{n})-\tilde{\Psi}(\bm{n}^{\bm{i},\bm{j}})}] = \Delta \Psi(\bm{n}) + O(\beta^3)
\end{align}
with
\begin{align}
    \Delta \Psi(\bm{n}) := \sum_{\langle \bm{i},\bm{j} \rangle} \delta_{\bm{i} \to \bm{j}}^{(\bm{n})} [\Psi(\bm{n}^{\bm{i},\bm{j}}) - \Psi(\bm{n})] .
\end{align}

We denote the right-hand side of Eq.~\eqref{eq:high_start} as $-\tilde{Q}(\bm{n})$:
\begin{widetext}
\begin{align}
    \tilde{Q}(\bm{n}) &= \sum_{\langle \bm{i},\bm{j} \rangle} [1-e^{\beta E(i_x-j_x)}] c(\bm{i} \to \bm{j}; \bm{n}) \nonumber \\
    &= \sum_{\langle \bm{i},\bm{j} \rangle} \delta_{\bm{i} \to \bm{j}}^{(\bm{n})} \left( -\beta E(i_x-j_x) + \beta^2 \left\{ \left(\frac{1}{2}+\tilde{\theta}_{\bm{i},\bm{j}}\right) E(i_x-j_x) [H(\bm{n}^{\bm{i},\bm{j}})-H(\bm{n})+E(i_x-j_x)] - \frac{1}{2} E^2 (i_x-j_x)^2 \right\} + O(\beta^3) \right) .
    \label{eq:high_Q}
\end{align}
\end{widetext}
% We now evaluate Eq.~\eqref{eq:high_Q} up to $O(\beta^2)$.
Here, the term proportional to $\beta$ vanishes since $\sum_{\langle \bm{i}, \bm{j} \rangle} \delta_{\bm{i} \to \bm{j}}^{(\bm{n})} (i_x-j_x) = 0$.
% \begin{align}
%     \tilde{Q}(\bm{n}) = - \beta \sum_{\langle \bm{i}, \bm{j} \rangle} \delta_{\bm{i} \to \bm{j}}^{(\bm{n})} E(i_x-j_x) = 0 .
% \end{align}
This confirms that the expansion of $\tilde{Q}(\bm{n})$ and thus that of $\tilde{\Psi}(\bm{n})$ start at $O(\beta^2)$, and leads to the Poisson-like equation for the lowest-order contribution:
\begin{align}
    \Delta \Psi(\bm{n}) = - Q(\bm{n}),
    \label{eq_poisson}
\end{align}
where $\tilde{Q}(\bm{n})=Q(\bm{n})+O(\beta^3)$.

Using $\delta_{\bm{i} \to \bm{j}}^{(\bm{n})} [H(\bm{n}^{\bm{i},\bm{j}})-H(\bm{n})] = n_{\bm{i}}(1-n_{\bm{j}}) U [\sum_{\bm{k}(|\bm{k}-\bm{i}|=1, \bm{k} \neq \bm{j})} n_{\bm{k}} - \sum_{\bm{k}(|\bm{k}-\bm{j}|=1,\bm{k} \neq \bm{i})} n_{\bm{k}}]$ in Eq.~\eqref{eq:high_Q}, we find
\begin{widetext}
% \begin{align}
%     \delta_{\bm{i} \to \bm{j}}^{(\bm{n})} [H(\bm{n}^{\bm{i},\bm{j}})-H(\bm{n})] = n_{\bm{i}}(1-n_{\bm{j}}) U \left[ \sum_{\bm{k}(|\bm{k}-\bm{i}|=1, \bm{k} \neq \bm{j})} n_{\bm{k}} - \sum_{\bm{k}(|\bm{k}-\bm{j}|=1,\bm{k} \neq \bm{i})} n_{\bm{k}} \right] ,
% \end{align}
\begin{align}
    Q(\bm{n}) &= \beta^2 \sum_{\langle \bm{i},\bm{j} \rangle} \delta_{\bm{i} \to \bm{j}}^{(\bm{n})} \left\{ \left(\frac{1}{2}+\tilde{\theta}_{\bm{i},\bm{j}}\right) E (i_x-j_x) [H(\bm{n}^{\bm{i},\bm{j}})-H(\bm{n})] + \tilde{\theta}_{\bm{i},\bm{j}} E^2(i_x-j_x)^2 \right\} \nonumber \\
    &= \beta^2 EU \sum_{\langle \bm{i},\bm{j} \rangle} \left(\frac{1}{2}+\tilde{\theta}_{\bm{i},\bm{j}}\right) n_{\bm{i}} (1-n_{\bm{j}}) \left[ \sum_{\bm{k}(|\bm{k}-\bm{i}|=1, \bm{k} \neq \bm{j})} n_{\bm{k}} - \sum_{\bm{k}(|\bm{k}-\bm{j}|=1, \bm{k} \neq \bm{i})} n_{\bm{k}} \right] (i_x-j_x) + \beta^2 E^2 \sum_{\langle \bm{i},\bm{j} \rangle} n_{\bm{i}} (1-n_{\bm{j}}) \tilde{\theta}_{\bm{i},\bm{j}} (i_x-j_x)^2 .
\end{align}
\end{widetext}
Thus, we can divide $Q(\bm{n})$ into two-body and three-body parts:
\begin{align}
    Q(\bm{n}) = \sum_{\bm{i},\bm{j}\in\Lambda} q_{\bm{i},\bm{j}}^{(2)} n_{\bm{i}} n_{\bm{j}} + \sum_{\bm{i},\bm{j},\bm{k}\in\Lambda} q_{\bm{i},\bm{j},\bm{k}}^{(3)} n_{\bm{i}} n_{\bm{j}} n_{\bm{k}} .
    \label{eq_decomp_charge}
\end{align}
% The coefficients $q_{\bm{i},\bm{j}}^{(2)}$ and $q_{\bm{i},\bm{j},\bm{k}}^{(3)}$ work as charges in the Poisson-like equation~\eqref{eq_poisson}.
Noticing $\sum_{\langle \bm{i},\bm{j} \rangle} n_{\bm{i}} (1-n_{\bm{j}}) \tilde{\theta}_{\bm{i},\bm{j}} (i_x-j_x)^2 = 0$,
% \begin{align}
%     \sum_{\langle \bm{i},\bm{j} \rangle} n_{\bm{i}} (1-n_{\bm{j}}) \tilde{\theta}_{\bm{i},\bm{j}} (i_x-j_x)^2 = 0,
% \end{align}
we finally obtain
\begin{align}
    q_{\bm{i},\bm{j}}^{(2)} &=
    \left\{
    \begin{aligned}
    &-\beta^2EU\theta & (\bm{i}-\bm{j}=\pm \bm{e}_x) \\
    &-2\beta^2EU\theta & (\bm{i}-\bm{j}=\pm \bm{e}_y) \\
    &\beta^2EU\theta & (\bm{i}-\bm{j}=\pm2\bm{e}_x, \bm{e}_x\pm\bm{e}_y, -\bm{e}_x\pm\bm{e}_y) \\
    &0 & (\text{otherwise})
    \end{aligned}
    \right.
    \label{eq_charge2}
\end{align}
and
\begin{align}
    q_{\bm{i},\bm{j},\bm{k}}^{(3)} &=
    \left\{
    \begin{aligned}
    &\beta^2EU & (\bm{j}=\bm{i}+\bm{e}_x, \bm{k}=\bm{i}\pm\bm{e}_y) \\
    &-\beta^2EU & (\bm{j}=\bm{i}-\bm{e}_x, \bm{k}=\bm{i}\pm\bm{e}_y) \\
    &0 & (\text{otherwise})
    \end{aligned}
    \right. .
    \label{eq_charge3}
\end{align}

In the same way as Eq.~\eqref{eq_decomp_charge}, by decomposing $\Psi(\bm{n})$ as
\begin{align}
    \Psi(\bm{n}) = \sum_{\bm{i},\bm{j}\in\Lambda} \psi_{\bm{i},\bm{j}}^{(2)} n_{\bm{i}} n_{\bm{j}} + \sum_{\bm{i},\bm{j},\bm{k}\in\Lambda} \psi_{\bm{i},\bm{j},\bm{k}}^{(3)} n_{\bm{i}} n_{\bm{j}} n_{\bm{k}} ,
\end{align}
Eq.~\eqref{eq_poisson} reduces to
\begin{align}
\left\{
\begin{aligned}
    & \Delta \psi_{\bm{i},\bm{j}}^{(2)} = - q_{\bm{i},\bm{j}}^{(2)} \\
    & \Delta \psi_{\bm{i},\bm{j},\bm{k}}^{(3)} = - q_{\bm{i},\bm{j},\bm{k}}^{(3)} ,
    \label{eq_poisson_red}
\end{aligned}
\right.
\end{align}
where the Laplacian is defined for $A \subset \Lambda$ as
\begin{align}
    \Delta \psi_A = \sum_{\langle \bm{i},\bm{j} \rangle, \bm{i} \in A, \bm{j} \notin A} (\psi_{A \backslash \{\bm{i}\}\cup\{\bm{j}\}} - \psi_A).
\end{align}
Here, $q_{\bm{i},\bm{j}}^{(2)}$ and $q_{\bm{i},\bm{j},\bm{k}}^{(3)}$ work as charges in the Poisson-like equations~\eqref{eq_poisson_red}.
This expression follows from the unique decomposition $\Delta \Psi(\bm{n}) = \sum_A (\Delta \psi_A) n^A$, where $\Psi(\bm{n}) = \sum_{A \subset \Lambda} \psi_A n^A$ and $n^A := \prod_{\bm{i} \in A} n_{\bm{i}}$~\cite{LefevereTasaki2005}.
% \begin{align}
%     \Delta \Psi(\bm{n}) = \sum_A (\Delta \psi_A) n^A
% \end{align}
% for
% \begin{align}
%     \Psi(\bm{n}) = \sum_{A \subset \Lambda} \psi_A n^A
% \end{align}
% with
% \begin{align}
%     n^A = \prod_{\bm{i} \in A} n_{\bm{i}} .
% \end{align}

According to Ref.~\cite{LefevereTasaki2005}, the obtained charge distributions [Eqs.~\eqref{eq_charge2} and \eqref{eq_charge3}] lead to the solutions $\psi_{\bm{i},\bm{j}}^{(2)} \sim 1/r^2$ for $E U \theta \neq 0$ and $\psi_{\bm{i},\bm{j},\bm{k}}^{(3)} \sim 1/r^5$ for $E U \neq 0$.
This means that, when $U > 0$ and $E > 0$ as considered in the main text, nonzero $\theta$ leads to long-range correlation at the leading order of the high-temperature expansion.

Note that the generalized exponential rate with $\theta\neq0$ does not satisfy the exponential condition in Ref.~\cite{Tasaki2004}:
\begin{align}
    \frac{\Phi(v,v';E)}{\Phi(v,v';-E)} = e^{\beta(2\theta U+E)} \neq e^{\beta E} ,
\end{align}
where $\Phi(v,v';E) := \phi(\beta(-v+v'-E))$.
% \begin{align}
%     \Phi(v,v';E) = \phi(-v+v'-E) .
% \end{align}

\section{Approximate fluctuating hydrodynamic equation}
\label{app_hydro}

To theoretically explain the $\theta$-dependent phase behavior, we derive an approximate fluctuating hydrodynamic equation [Eq.~\eqref{eq_linear_fluct_hydro}] from the lattice gas model following the path-integral formalism~\cite{Andreanov2006, Lefevre2007}.

\subsection{Path-integral formulation and Langevin equation}

We start by describing the time evolution of the system using the path probability density.
Introducing time-dependent auxiliary fields $\tilde{\bm{n}}(t) := \{ \tilde{n}_{\bm{i}}(t) \}_{\bm{i}}$, the path probability can be cast into the form $P[\bm{n}] \propto \int D\tilde{\bm{n}} \exp(-S[\bm{n}, \tilde{\bm{n}}])$~\cite{Andreanov2006, Lefevre2007}.
The action $S[\bm{n}, \tilde{\bm{n}}]$ is given by
\begin{align}
    S[\bm{n}, \tilde{\bm{n}}] & = \int dt \sum_{\bm{i}} \bigg[ -i \tilde{n}_{\bm{i}} \partial_t n_{\bm{i}} \nonumber \\
    & \quad - \sum_{\bm{j} \, (\sim \bm{i})} w_{\bm{i}\to \bm{j}}(\bm{n}) n_{\bm{i}} (1-n_{\bm{j}}) \left(e^{i(\tilde{n}_{\bm{i}} - \tilde{n}_{\bm{j}})} - 1\right) \bigg],
\end{align}
where $\sum_{\bm{j} \, (\sim \bm{i})}$ denotes the sum over all sites adjacent to site $\bm{i}$.

To obtain a tractable continuous equation, we apply the Gaussian approximation to the auxiliary field $\tilde{n}_{\bm{i}}$.
By expanding the exponential term up to the second order, $e^{i(\tilde{n}_{\bm{i}} - \tilde{n}_{\bm{j}})} - 1 \simeq i(\tilde{n}_{\bm{i}} - \tilde{n}_{\bm{j}}) - (\tilde{n}_{\bm{i}} - \tilde{n}_{\bm{j}})^2 / 2$, the action $S[\bm{n}, \tilde{\bm{n}}]$ becomes quadratic in $\tilde{\bm{n}}$.
The path probability density can then be written as
\begin{align}
    P[\bm{n}] & \propto \int D\tilde{\bm{n}} \exp\bigg[ -i \int dt \sum_{\bm{i}} \tilde{n}_{\bm{i}} \left( \partial_t n_{\bm{i}} - F_{\bm{i}}(\bm{n}) \right) \nonumber \\
    & \quad - \frac{1}{2} \int dt \sum_{\bm{i},\bm{j}} \tilde{n}_{\bm{i}} M_{\bm{i}\bm{j}}(\bm{n}) \tilde{n}_{\bm{j}} \bigg],
\end{align}
where $F_{\bm{i}}(\bm{n}) := -\sum_{\bm{j} \, (\sim \bm{i})} [w_{\bm{i}\to \bm{j}}(\bm{n}) n_{\bm{i}} (1-n_{\bm{j}}) - w_{\bm{j}\to \bm{i}}(\bm{n}) n_{\bm{j}} (1-n_{\bm{i}})]$, and $M_{\bm{i}\bm{j}}(\bm{n})$ is a symmetric matrix.

We introduce noise fields $\bm{\xi}(t) := \{ \xi_{\bm{i}}(t) \}_{\bm{i}}$ as
\begin{align}
    & \exp\bigg( -\frac{1}{2} \int dt \sum_{\bm{i},\bm{j}} \tilde{n}_{\bm{i}} M_{\bm{i}\bm{j}}(\bm{n}) \tilde{n}_{\bm{j}} \bigg) \nonumber \\
    & \propto \int D \bm{\xi} \exp\bigg( -\frac{1}{2} \int dt \sum_{\bm{i},\bm{j}} \xi_{\bm{i}} M_{\bm{i}\bm{j}}^{-1}(\bm{n}) \xi_{\bm{j}} + i \int dt \sum_{\bm{i}} \tilde{n}_{\bm{i}} \xi_{\bm{i}} \bigg).
\end{align}
Substituting this identity back into the path integral, the integration over $\tilde{\bm{n}}$ yields a product of delta functionals:
\begin{align}
    P[\bm{n}] & \propto \int D\xi \prod_{\bm{i}} \delta\left( \partial_t n_{\bm{i}} - F_{\bm{i}}(\bm{n}) - \xi_{\bm{i}} \right) \nonumber \\
    & \quad \times \exp\bigg( -\frac{1}{2} \int dt \sum_{\bm{i},\bm{j}} \xi_{\bm{i}} M_{\bm{i}\bm{j}}^{-1}(\bm{n}) \xi_{\bm{j}} \bigg).
\end{align}
This expression demonstrates that the configuration $\bm{n} (t)$ follows a Langevin equation:
\begin{equation}
    \partial_t n_{\bm{i}} = F_{\bm{i}}(\bm{n}) + \xi_{\bm{i}}(t),
\end{equation}
where $\xi_{\bm{i}}(t)$ is a zero-mean Gaussian white noise satisfying $\langle \xi_{\bm{i}}(t)\xi_{\bm{j}}(t') \rangle = M_{\bm{i}\bm{j}}(\bm{n}) \delta(t-t')$.
The covariance matrix elements are given by $M_{\bm{i}\bm{i}}(\bm{n}) = \sum_{\bm{j} \, (\sim \bm{i})} [w_{\bm{i}\to \bm{j}} n_{\bm{i}} (1-n_{\bm{j}}) + w_{\bm{j}\to \bm{i}} n_{\bm{j}} (1-n_{\bm{i}})]$, $M_{\bm{i}\bm{j}}(\bm{n}) = - [w_{\bm{i}\to \bm{j}} n_{\bm{i}} (1-n_{\bm{j}}) + w_{\bm{j}\to \bm{i}} n_{\bm{j}} (1-n_{\bm{i}})]$ for adjacent pairs $(\bm{i}, \bm{j})$, and $M_{\bm{i}\bm{j}}(\bm{n}) = 0$ otherwise.

\subsection{Continuum limit and linearization}

We focus on the density fluctuations around the homogeneous state with a mean density $\rho_0 = 1/2$.
Letting $n_{\bm{i}}(t) = \rho_0 + \phi(\bm{r}_{\bm{i}}, t)$ where $\phi$ is small, we linearize the transition rates $w_{\bm{i}\to \bm{j}}(\bm{n})$ with respect to $\phi$ and keep terms up to $O(\phi)$.
The noise covariance $M_{\bm{i}\bm{j}}(\bm{n})$ is also approximated by its value at the mean density $\rho_0$, making the noise additive.
Transforming the linearized Langevin equation into the Fourier space, we obtain
\begin{equation}
    \partial_t \phi_{\bm{k}} = - A_{\bm{k}} \phi_{\bm{k}} + \xi_{\bm{k}},
    \label{appeq_linear_hydro_ft}
\end{equation}
where $\langle \xi_{\bm{k}}(t) \xi_{\bm{k}'}^*(t') \rangle = 2 L^2 B_{\bm{k}} \delta_{\bm{k}, \bm{k}'} \delta(t-t')$.

By expanding the coefficients for small wave vectors $\bm{k} = (k_x, k_y)$ up to $O(k^2)$, we identify the deterministic and noise terms corresponding to Eq.~\eqref{eq_linear_fluct_hydro} in the main text.
Specifically, $A_{\bm{k}}$ is given by $A_{\bm{k}} \simeq A_x {k_x}^2 + A_y {k_y}^2$, and the noise intensity is $B_{\bm{k}} \simeq B_x {k_x}^2 + B_y {k_y}^2$, with the following coefficients:
\begin{equation}
\left\{
\begin{array}{l}
    A_x = e^{\theta E} \cosh(E/2) \left\{1 - (5 U / 4) \left[1 + 2\theta \tanh(E/2)\right]\right\}, \\
    A_y = 1 - 5 U / 4, \\
    B_x = (1/4) e^{\theta E} \cosh(E/2), \\
    B_y = 1/4.
    \label{appeq_hydro_coeffs}
\end{array}
\right.
\end{equation}
Under these approximations, Eq.~\eqref{appeq_linear_hydro_ft} is expressed in the real space as
\begin{equation}
    \partial_t \phi = A_x {\partial_x}^2 \phi + A_y {\partial_y}^2 \phi + \xi,
\end{equation}
with $\braket{\xi (\bm{r}, t)} = 0$ and $\braket{\xi (\bm{r}, t) \xi (\bm{r}', t')} = -2 (B_x {\partial_x}^2 + B_y {\partial_y}^2) \delta (\bm{r} - \bm{r}') \delta (t - t')$, which is equivalent to Eq.~\eqref{eq_linear_fluct_hydro}.

The steady-state structure factor in the homogeneous phase is analytically evaluated as $S_{\mathrm{hy}}(\bm{k}) = \lim_{t\to\infty} \langle |\phi_{\bm{k}}(t)|^2 \rangle / L^2 = B_{\bm{k}} / A_{\bm{k}}$.
The discontinuity of the structure factor at the origin, which signifies the presence of long-range correlations, is given as
\begin{equation}
    \Delta S_{\mathrm{hy}} = \lim_{k_x \to 0} \lim_{k_y \to 0} S_{\mathrm{hy}}(\bm{k}) - \lim_{k_y \to 0} \lim_{k_x \to 0} S_{\mathrm{hy}}(\bm{k}) = B_x / A_x - B_y / A_y.
\end{equation}
Substituting Eq.~\eqref{appeq_hydro_coeffs} yields Eq.~\eqref{eq_hydro_structure_factor} of the main text.

% \bibliography{ref.bib}
\input{output.bbl}

\end{document}